# How to organize an online conference

Orad Reshef[1], Igor Aharonovich[2], Andrea Armani[3], Sylvain Gigan[4], Rachel Grange[5], Mikhail A. Kats[6], Riccardo Sapienza[7]

[1]University of Ottawa, Canada, [2]University of Technology, Australia, [3]University of Southern California, USA, [4]Sorbonne Universite, France, [5]ETH Zurich, Switzerland, [6]University of Wisconsin-Madison, USA, [7]Imperial College London, UK

**Abstract:** On January 13th 2020, the inaugural Photonics Online Meetup (POM) brought together more than 1100 researchers to discuss the latest advances in photonics. Or rather, it didn't, because the meeting was completely delocalized with the speakers, organizers, and attendees scattered across six continents and hundreds of locations, connected via a video-conferencing tool and social media. Despite this "delocalisation", the meeting retained many characteristics of a traditional conference: invited and contributed talks with follow-up questions and discussion, and a poster session. However, unlike with traditional conferences, all attendees avoided air travel, high registration costs (it was completely free), CO2 emission, or visa issues. Surely the impact on families was minimized as well, though participants in inconvenient time zones had to wake up early or stay up late. In this Comment, we highlight the key steps that enabled this event, offering tips and advice, to aid the organisation of similar Online Meetups.

The format and venues of academic and scientific conferences have not meaningfully changed in centuries. In fact, conferences today closely resemble the meetings of the Royal Society or ancient Greece from centuries ago. Therefore, it is not surprising that many in the academic community are hesitant to deviate from this technique despite the emergence of many web-based alternatives.

However, with the rising impact of academic travel on the environment and on work–life balance, and the increasing availability of fast and reliable internet connections, there is now room for a paradigm shift enabled by modern teleconference solutions. To this end, several of us from around the world assembled to organise the first online-only conference focused on innovations in optics, forming the *Photonics Online Meetup (POM)*. Our vision was to provide a free, globally accessible meeting where neither the speakers nor the participants needed to be co-located. In addition to disseminating technical information, an online meeting can help reduce the carbon footprint of conference travel, reduce the burden on families, in particular time spent away from home, eliminate the cost of conference participation, and improve access to high-level technical content facilitating attendance to all. At the time of writing, APS March Meeting 2020 was just cancelled due to the outbreak of the coronavirus disease 2019 (COVID-19); we believe that online meetings can be particularly beneficial at such times. A personal view of the event from two participants can be found in refs [1,2] while from the organiser in refs. [3,4].

## The format

The format of POM was deliberately chosen to be very similar to that of conventional conferences. We wanted the event to be accessible in many time zones, therefore we chose a relatively short format of 3 sessions, 1.5 hours each, separated by short breaks, for a duration of 5 hours in total. We selected January 13, 2020, for the inaugural event to try to avoid the winter holidays and to align with the academic calendar at most universities. Complementary to the live event, we also planned a poster session, held on Twitter starting January 9, 2020. The talks were recorded but available online only for two weeks, to keep the immediacy and spirit of conventional conferences, and to provide speakers the opportunity to present unpublished results without fear of permanent public disclosure.

## The team

Key to every event—especially so for a large-scale meeting—is a motivated and organized team. Our team assembled quickly following a discussion on Twitter, with seven academics working in diverse areas of photonics, and located on different continents. As it was the first event of its kind, the overall topic of the conference—Photonics—emerged naturally. Orad Reshef and Andrea Armani took the reins as chairs, and we collectively worked on planning the event, assembling the program, inviting the speakers, and running the event. Due to the different time zones, a quick turnaround and an enthusiastic engagement have been crucial: choose your organising team judiciously!

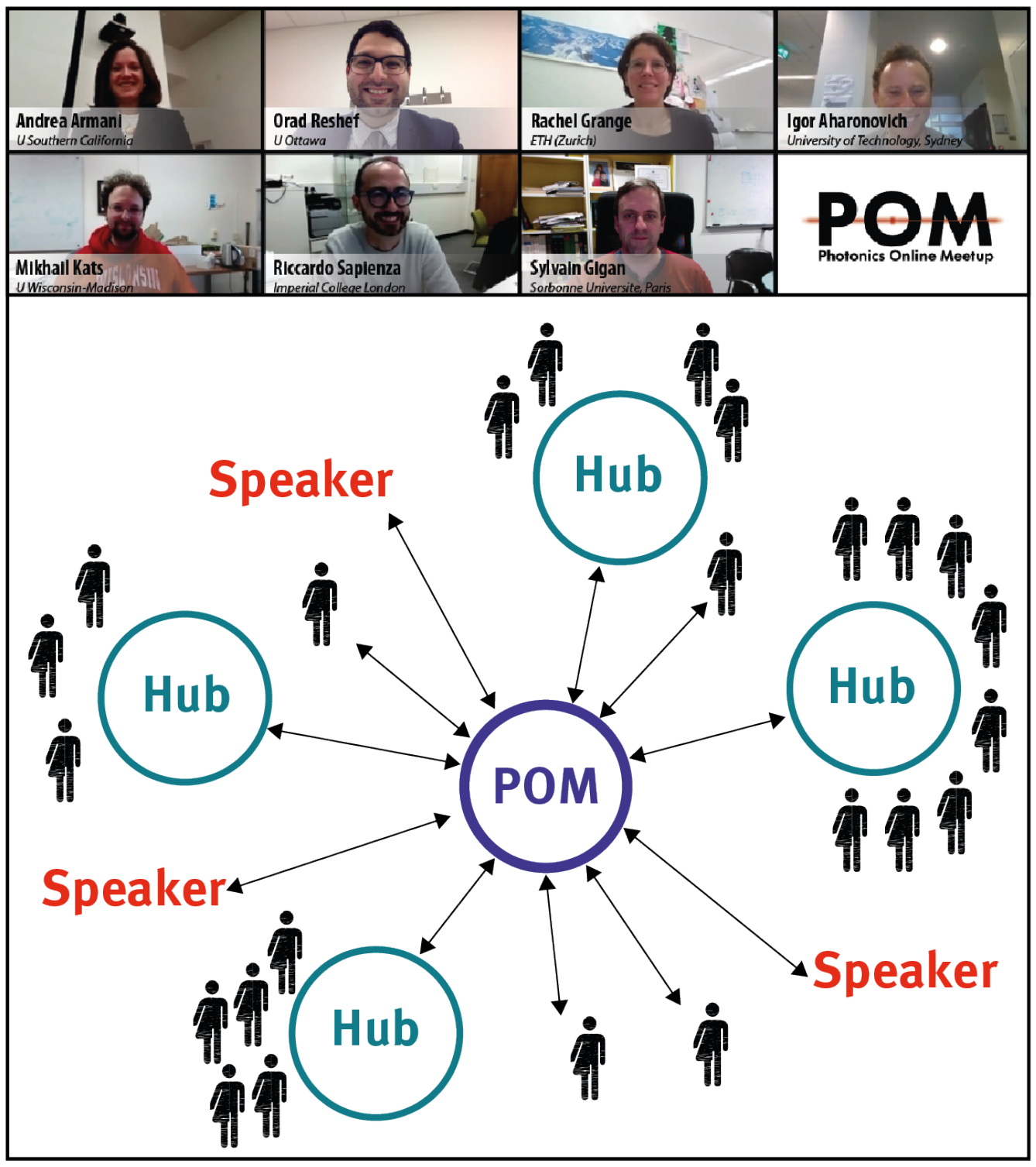

Top: *the organizing committee.* Bottom: *Schematic showing interconnectivity of POM.*

## The infrastructure

*Communication*. Since our team was scattered around three continents and multiple time zones, we took advantage of asynchronous messaging/work tools to organize the event. In particular, we chose Slack as our communication platform as many of us already use it for research and other purposes. We managed to work around the clock, e.g., with our Australian co-organizer picking up when the Americans were ending the day, and the other way around.

*Video platform.* We enlisted the help of IT support at the University of Southern California (USC) to accommodate multiple speakers from so many locations and to facilitate a connection between a large number of participants while minimizing technical risks. Based on their guidance, WebEx Events was chosen as it can simultaneously connect hundreds of participants globally with additional controls, such as audience muting and delocalized presenters. We believe that Zoom and other similar software can be used as well. USC also provided technical support in ahead of and throughout the event. In this manner, USC acted as a centralized control hub, with all speakers and participants connecting to the primary site. We believe that real-time professional technical support at a central control hub is essential to a successful online meeting.

*Website.* The last basic infrastructure piece was a website for researchers to submit abstracts and register for the event. The website acted like a central repository for all information, was built by the chair using WordPress, and was hosted by USC. Abstract submissions and event registrations were done using a simple survey software. Notably, because the event was free, secure information like credit card numbers did not need to be transmitted. This aspect greatly simplified the registration process.

*Advertisement*. The POM was organised in less than 3 months, and most of the advertisement was done through Twitter, which is the most common platform for academic interaction. Because of the new conference format and the lack of a sponsoring professional society, engaging the scientific community was critical to the event's success. The call for abstracts was then promoted via various social media, including Twitter, Facebook, LinkedIn, and Instagram, as well as emails to colleagues. We encouraged submission of abstracts for either oral presentations or online posters. Social media acted as a seed, and then word-of-mouth helped us reach a large audience. Eventually we had more than 1000 attendees and a peak of 600 simultaneous connections.

**The program**

The first step in building the program was selecting the three topics for POM. As a new event, we had a lot of flexibility, so the initial list of proposed topics ranged from fundamental optics topics to various applied technologies. The committee voted, and the top-ranked themes were nanoscale quantum optics, optical materials, and integrated optics.

For each of the three sessions, we formed smaller topic committees that identified potential invited speakers. Unlike a conventional conference, often suffering from moderate invited speaker acceptance rates, every speaker who we invited accepted enthusiastically (perhaps excited by the opportunity to speak without boarding an airplane!)

By the deadline, we received about 100 abstracts coming from PhD students, postdoctoral researchers, and faculty members. From these applications, our subcommittees selected 9 for oral presentations (3 per track), and the others were accepted as poster presentations.

**The Hubs**

We were initially quite concerned that the online format would result in fewer opportunities for networking, brainstorming, and other interactions. To that end, we encouraged the creation and registration of local viewing sites at various institutions, which we called ”POM-Hubs”. We anticipated that a POM-Hub would be hosted in a conference room or auditorium at an institute or university, where POM participants could gather into a more “social”, in-person event.

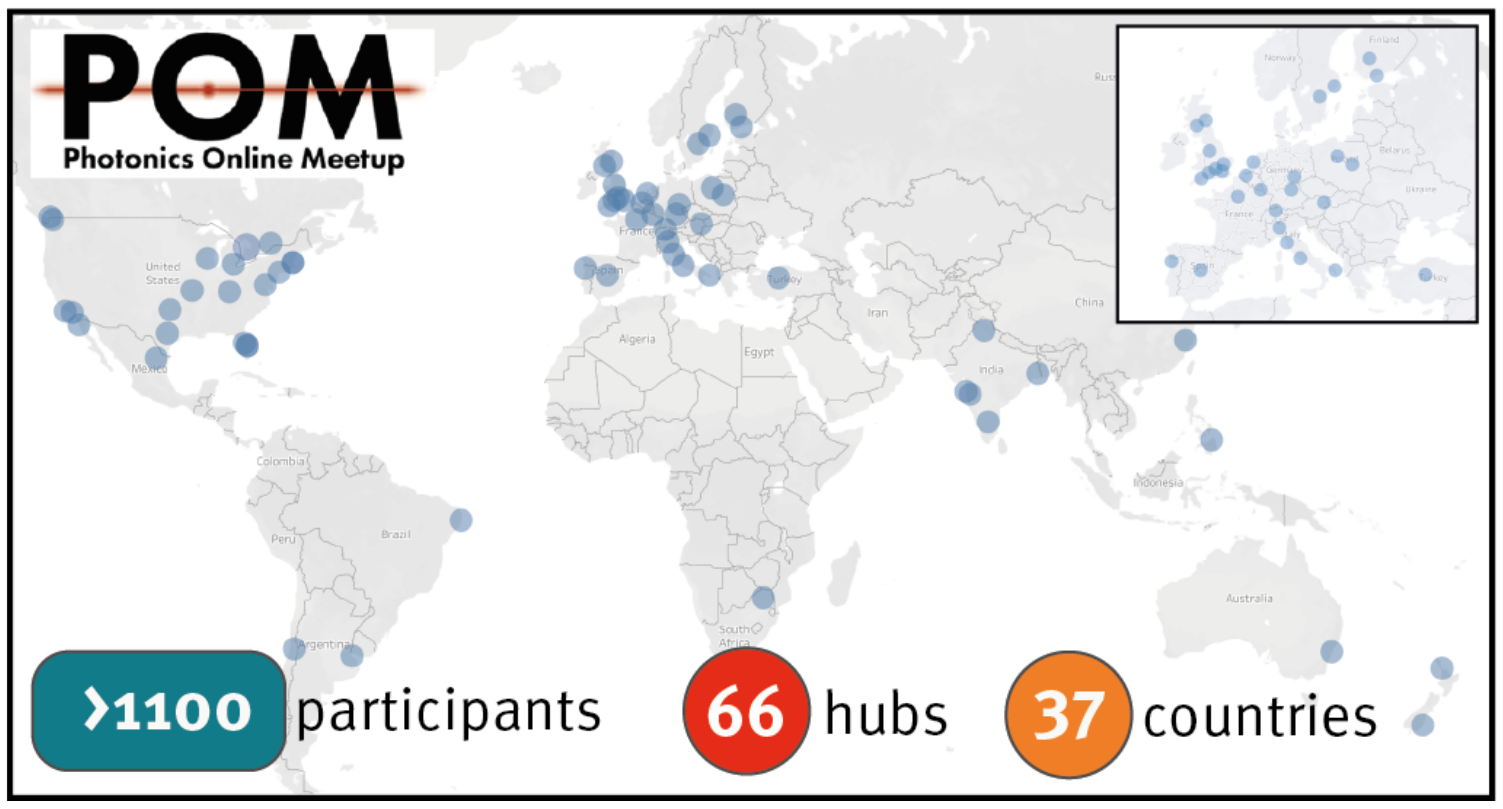

*Map of the POM hubs. Inset: Zoom-in of the hubs across Europe.*

We promoted the Hubs on our website and encouraged the organizers to actively promote their hub and the conference locally. Ultimately, 66 hubs located at universities as well as other scientific institutions, such as OSA, SPIE, and Nature, emerged on all continents (except Antarctica). In many places, the hubs were organized and supported by local OSA (The Optical Society) and SPIE student chapters. The hubs were very much like mini in-person conferences: groups of students, postdocs, and senior researchers watched the talks together, many of the hubs organized snacks, poster sessions, and there were plenty of opportunities for informal interaction. Despite the short (5-hour) duration of POM, some conference culture began to emerge: a hub at UC-Davis arranged a POM bingo game (see figure), which then made its way to other participants via social media.

**POM Bingo – UC Davis Hub**

| | | | | |
|---|---|---|---|---|
| **Applause after a talk** | **Exciton** | **Neural network** | **Single photon emission** | **Hanbury Brown-Twiss** |
| **Teleportation** | **Refraction** | **Cold atoms** | **Integrated** | **Technical issues** |
| **Nanoparticle** | **Cat enters screen** | **Waveguide** | **Quantum dot** | **Terahertz** |
| **Microwaves** | **Topology** | **Hong-Ou-Mandel** | **Tweet local hub photo** | **Lunch arrives** |
| **Photonics** | **Polymer** | **Projector overheats** | **Online question asked** | **Metamaterial** |

*The #POM bingo board, made on the fly during the event by the UC-Davis hub (courtesy of Marina Radulaski)*

We were amazed that even in India and Australia, where the meeting started in the middle of the night and at 6 am, respectively, the hubs ran at full steam.

## The poster session

Given the limited number of speaking slots and the high number and quality of the abstract submissions, we wanted to increase the opportunities for researchers to disseminate their findings. This demand motivated the idea of a Twitter poster session.

We provided a four-slide poster template that was optimized for display on Twitter. We asked the presenters to create a personal or group Twitter account (if they didn't already have one), post their poster, and add a short description. We also encouraged the use of the conference hashtag (#POM20) to allow researchers to easily join the poster session. For those without an account, such as researchers in locations where access to Twitter is restricted, the organizers posted the posters via the POM twitter account and relayed the questions to the presenters directly. Comments and answers could be followed by all as a thread below each poster.

As compared to conventional conferences, the poster on twitter had a much wider reach, with some reaching 4000 views. While this is not a typical "wine and cheese" poster viewing, the posters were viewable for days, and one can still go back to see them by looking up the hashtag. This method also allowed researchers to create threads, directly linking relevant papers to their posters, further improving research dissemination.

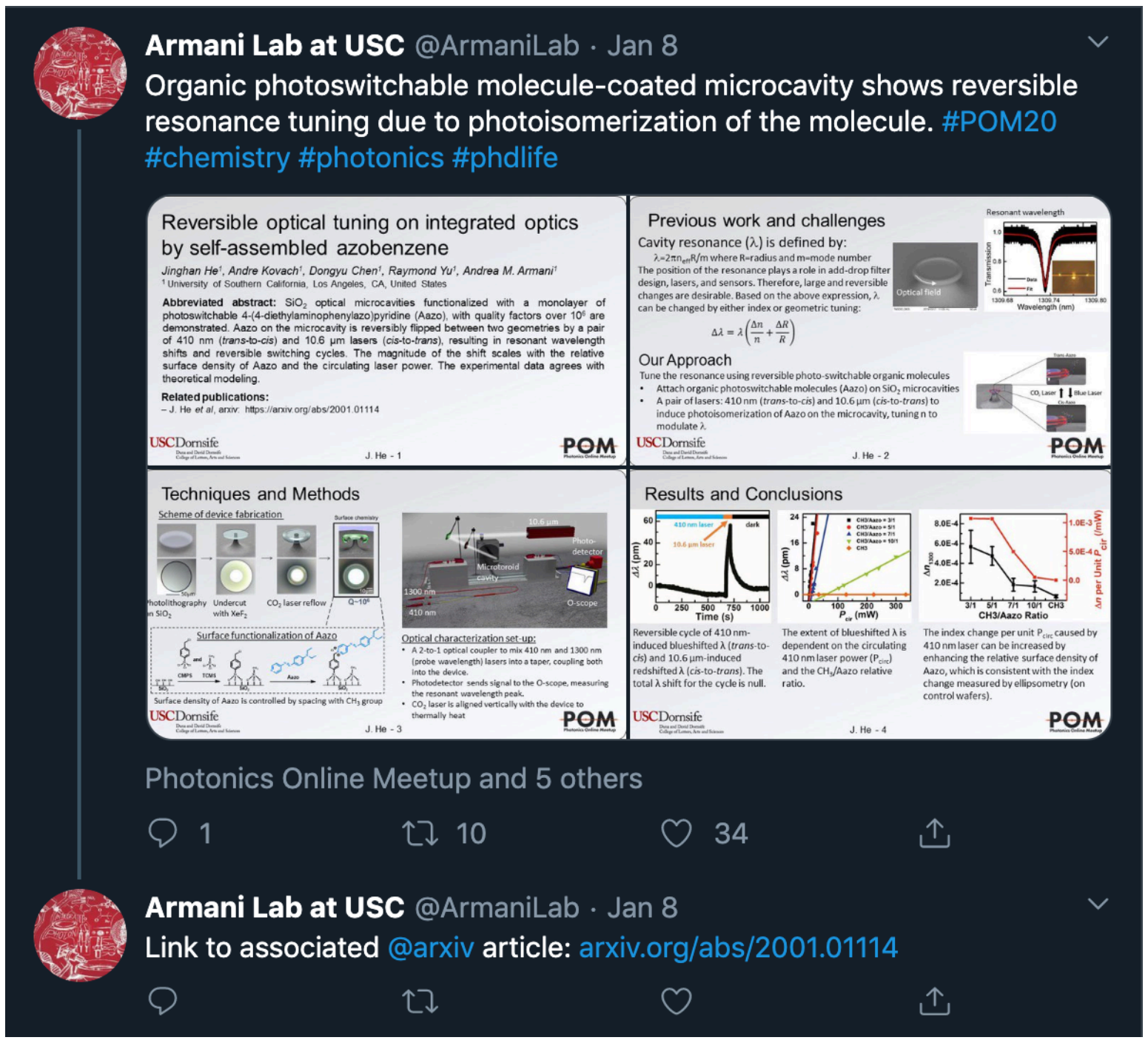

*An example of a Twitter-based poster hosted during the virtual poster session.*

## Before the day of the conference

An important aspect of the preparation was to make sure that the conference would not be derailed by technical problems. It was therefore very important to brief the presenters on the software, to check the quality of the video and sound, and to generally minimize IT issues. Each speaker performed a test run with the IT support team at USC. We also used the website to post test links for the software for individuals and Hubs in advance and a "helpful hints" document with instructions on using the software and troubleshooting.

**During POM**

The videoconference was accessible from all hubs and registered participants a few minutes before the conference. Two staff from USC were also participating and ready to step in for technical issues (which happened with one speaker). All the POM organizers, usually present at their own local hubs, kept in touch via Slack.

The chairs started each session and had a control over the Webex software to mute/intervene if necessary. While on-screen, the chairs broadcast their videos from a private office to reduce the noise from their hubs.

We asked hub participants to Tweet photos from their hubs, with the hashtag of the conference. This activity triggered a tremendous amount of interaction online between the hubs throughout the whole conference. Depending on the time of the day, several hubs had organized food (e.g., coffees in the morning, pizza in the evening…) and we were thrilled to witness that the attendance of the hubs varied from a handful to more than 50 for some, despite being extremely early in the morning or in the middle of the night. We tried to promote those pictures of the hubs during the break, which definitely created a feeling of community, both in and between the different participants.

We asked all speakers to be online and ready to present on the videoconferencing software at the start of the conference. This way, in the event of a technical glitch, we could immediately proceed to the following speaker. One such issue did occur, where a speaker could not connect, and it was resolved by continuing onto the next talk and moving that speaker to the next session.

Questions from the audience were typed into Webex chat, asked by the session chairs, and answered by the speakers. As will be discussed, this aspect proved to be a challenge, though we can now propose some basic solutions based on our experience.

## POM20: ANALYSIS AND PERSPECTIVES

In evaluating the event's success, we considered its global reach, particularly into communities that might not normally attend conventional conferences, as well as absolute participant numbers. We also surveyed the participants before and after the event to learn as much as possible about their experience.

Overall, we evaluate this inaugural Photonics Online Meetup to be a strong success. With 66 hub sites in 27 countries on 6 continents, the event had a large impact on an international scale. More than 600 participants watched the event at the Hub sites, with approximately 500 more participating as individuals. These viewers were able to interact with the presenters through the Q&A panel, with the Session Chairs merging questions and providing clarification to enable a smooth Q&A session. The recorded videos reached an additional 200 unique viewers. Importantly, more than half of the participants were graduate or undergraduate students. For many of these students, POM was their first conference, and it would not have been possible if it had not been free.

We would like to emphasize some of the innovations of POM as compared to in-person conferences and existing online events such as webinars. Our virtual poster session was held entirely on Twitter via a specially designed poster format. It included the retweeting of posters via the conference account, and almost 60 high-quality posters were made available to everyone online, with an average of 3,000+ impressions recorded per poster in the first three days -- far more than one would expect at an in-person conference. Our POM-Hub model successfully merged talks broadcasted over the internet with a physical presence and community building: the 66 POM-Hubs organized organically around the world ranged from a few students getting together for pizza and POM presentations in Stockholm to a massive 65-person event in Ottawa. One Hub even developed a POM bingo game, which spread in real time via Twitter. We were thrilled to see photos in real time of students at a POM-Hub in Bombay, India at 3am, engaged in the presentations.

### Challenges

Despite the success of this initial event, the post-event surveys from participants revealed several challenges.

The most interesting is probably related to the emotional engagement due to the lack of in-person interaction. It is still an open question how to provide a fuller experience with online networking. For example, the lack of clapping after each talk was distinctly felt as something missing. In the future, chat rooms could be used to stimulate small-group discussion, and help the networking aspect; even virtual-reality technologies could be explored, though this would require additional equipment and expertise.

Like with most online events, audio and video quality was the primary issue raised by participants. Despite having every speaker test the software in advance, we learnt during the conference that several hubs had audio issues, often due to connectivity issues at the hub site or poor audio quality from the speakers' microphones. These types of challenges can be resolved at least partially by testing the connection at all of the hubs in advance and using external microphones, which can improve the audio clarity.

Interestingly, the participants had mixed views on the conference size and the selection of topics. While many participants appreciated the smaller scale, as it allowed them to attend all of the talks, other participants wished for more presentations on a wider range of topics. Additionally, given the diverse educational level, some respondents suggested more in-depth tutorial-type presentations and extended Q&A periods; this was also pointed out by David Pile in a post-conference debrief [2]. One of the invited speakers, Nader Engheta, suggested that it may be better to have questions come directly from the listeners via an audio/video connect, rather than via the session chairs [2].

One approach to addressing some of these suggestions is to have a pair of events: one consisting of tutorial-type lectures and a second consisting of more detailed technical presentations. Another option is parallel streaming. However, having multiple virtual rooms would require the different hubs to select a single topic. At our inaugural POM, we decided to avoid this type of fragmentation.

In addition, one of the main goals of the event was to improve access to scientific findings and to increase education equality. For this goal to be realized, everyone on the global stage must have access to the software and various web platforms used. Despite an exhaustive search, we were unable to find universally accessible web platforms that met our needs for both the poster session and the online presentations. As a result, some regions were partially excluded from the event due to time-zone challenges or the exclusion of certain users from social media. In the future, this type of access is a fundamental challenge facing not only the scientific community, but the global community, that will require multi-stakeholder engagement to overcome.

## The vision ahead

Given the clear benefits to multiple research communities and the cost (both financial and otherwise) of holding in-person events, we aspire to establish a movement of online meetups. The suite of freely (or almost free) technologies that is now available was not accessible even a decade ago, and we should take advantage of it. For some scientists around the world, particularly students, POM was a truly unique opportunity to experience an academic meeting.

As suggested by Miles Padgett, pre-prepared subtitles may make these types of talks even more accessible [5]. Furthermore, as novel technology is developed, e.g. virtual reality, online meetings will likely become even more compelling.

## Conclusions

We evaluate POM20 to be largely a success, generating excitement and enabling the sharing of scientific knowledge in a more-inclusive and cost-effective way.

It is important to note that before this event, none of the organizers had worked together, and we assembled around a common shared vision. We sincerely hope that the participants enjoyed POM20, and we thank them for their time and commitment.

We believe that the innovations we developed in preparation for the inaugural POM event will be applicable not only to future POMs, but can be readily generalized and implemented for low-cost, inclusive and delocalized conferences in all fields. Though we do not expect Online Meetups to displace all physical conferences, nor to replace the serendipity and networking of meeting in person, perhaps online events can encourage a reduction in air travel and reduce the significant costs of conference attendance and impact to families for scientists and other academics as well as democratizing access to knowledge on a global scale.

### *Acknowledgement*

We thank Gigi Ragusa and Antonia Sladek for post-event interviews and feedback. We once again thank the USC Technical team for their invaluable support and all the volunteers that assisted us and inspired us along the way. #POM21

### *Conflict of Interest*

This effort received no financial support from the social-media and/or networking-software companies mentioned, including Slack, Twitter, and WebEx. Thus, there are no conflicts of interest to report.

### References:

[1] Virtual conferences get real *(Nature Reviews Materials (2020))*

*[2]* Photonics from afar *(Nature Photonics 14, 137(2020))*

*[3]* Rethinking conferences *(Times Higher Education)*

*[4] A Different Kind of Conference (Optics & Photonics News)*

*[5] https://twitter.com/milespadgett/status/1233470052646768641?s=21*